\documentclass[manuscript]{acmart}

\AtBeginDocument{%
  }

\setcopyright{acmlicensed}
\copyrightyear{2018}
\acmYear{2018}
\acmDOI{XXXXXXX.XXXXXXX}

\acmConference[Conference acronym 'XX]{Make sure to enter the correct
  conference title from your rights confirmation emai}{June 03--05,
  2018}{Woodstock, NY}
\acmISBN{978-1-4503-XXXX-X/18/06}

\begin{document}

\title{Enhancing Well-Being Through Positive Technology: VR Forest Bathing}


\author{Francisco R. Ortega}
\affiliation{%
  \institution{Colorado State University}
  \city{Fort Collins}
  \state{CO}
  \country{USA}
}
\orcid{0002-2449-3802}
\email{fortega@colostate.edu}

\author{Victoria Interrante}
\affiliation{%
  \institution{University of Minnesota}
  \city{Minneapolis}
  \state{MN}
  \country{USA}
}
\orcid{0002-3313-6663}
\email{interran@umn.edu}

\author{Sara LoTemplio}

\affiliation{%
  \institution{Colorado State University}
  \city{Fort Collins}
  \state{CO}
  \country{USA}
}
\orcid{0003-0142-3702}
\email{sara.lotemplio@colostate.edu}

\author{Rachel Masters}
\affiliation{%
  \institution{Colorado State University}
  \city{Fort Collins}
  \state{CO}
  \country{USA}
}
\orcid{0002-3857-6537}
\email{Rachel.Masters@colostate.edu}

\author{Jalynn Nicoly}
\affiliation{%
  \institution{University of Colorado}
  \city{Boulder}
  \state{CO}
  \country{USA}
}
\orcid{0002-2897-5833}
\email{Jalynn.Nicoly@colorado.edu}

\author{Zahra Borhani}
\affiliation{%
  \institution{Colorado State University}
  \city{Fort Collins}
  \state{CO}
  \country{USA}
}
\orcid{0003-4242-551X}
\email{zahra.borhani@colostate.edu}

\author{Deana Davalos}
\affiliation{%
  \institution{Colorado State University}
  \city{Fort Collins}
  \state{CO}
  \country{USA}
}
\orcid{0003-2727-3634}
\email{Deana.Davalos@colostate.edu}

\author{Daniel Zielasko}

\affiliation{%
  \institution{Trier University}
  \city{Trier}
  \state{}
  \country{Germany}
}
\orcid{0003-3451-4977}
\email{daniel.zielasko@rwth-aachen.de}
\renewcommand{\shortauthors}{Interrante et al.}

\begin{abstract}
The growing demand for accessible therapeutic options has led to the exploration of Virtual Reality (VR) as a platform for forest bathing, which aims to reduce stress and improve cognitive functions. This paper brings together findings from three studies by the authors. The first study compared environments with and without plant life to examine how biomass influences stress reduction. The second study focused on the differences between low-fidelity and high-fidelity VR environments, while the third explored whether the benefits of VR forest bathing come from being immersed in realistic environments or simply from viewing something beautiful. The results showed no significant differences between environments with and without biomass, but highlighted the positive effects of high-fidelity VR environments and realistic nature over abstract art. The paper also covers how VR nature experiences may boost executive functioning and well-being in older adults and discusses the potential of generative AI to create customized VR environments. It concludes with a call for further collaborative research to refine VR forest bathing for stress relief and cognitive enhancement.
\end{abstract}


\begin{teaserfigure}
  \includegraphics[width=\textwidth]{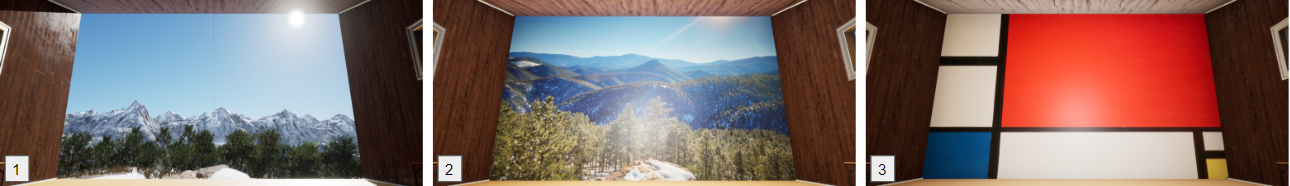}
  \caption{3D Moving Forest, Real Forest Image, and Abstract Art Environments from ~\cite{jnicoly2024sap}.} 
  \Description{Three experimental VR environments (3D Moving Forest, Real Forest Image, and Abstract Art Environment) from~\cite{jnicoly2024sap}}
  \label{fig:teaser}
\end{teaserfigure}

\received{20 February 2007}
\received[revised]{12 March 2009}
\received[accepted]{5 June 2009}

\maketitle

\section{Introduction}

Large trees and Superior Lake surround Duluth, Minnesota (USA). Spending time in such a location can have an impact on a person. The positive effects of nature, particularly a forest, have been shown to reduce stress~\cite{yao2021effect, annerstedt2010finding} and increase executive functions~\cite{cai2024effects}. The drive from Duluth to Minneapolis, MN, is so immersive because of its nature that one ponders the privilege of being in such an experience and its benefits. 

Shinrin-yoku (\textbf{forest bathing}) was coined by the Japanese Ministry of Agriculture, Forestry, and Fisheries in 1982~\cite{park2010physiological}. This practice involves people spending time in nature, which has been shown~\cite{kaplan1995restorative,ulrich1981natural} to reduce stress and well-being in general~\cite{10665-345751}. However, many people have no access to nature due to their typical day-to-day urban life routines, mobility problems, or other impediments, which deprive them of the benefits of forest bathing~\cite{soga2016extinction}. It would be necessary to see how much stress reduction and improvement of cognitive functions users can gain when accessing a simulated natural space, making virtual reality (VR) a great equalizer. 

Another important aspect to discuss in this context is the~\textit{Biophilia Hypothesis}, which states that people will feel more restored after being immersed in a place rich in Biomass (i.e., living biological organisms) compared to urban areas lacking biomass, all other factors being equal. Yet, in a VR forest bathing experience, nothing is alive, so understanding how virtual elements play a role is critical for the development of this positive technology. We have been able to answer a few questions with our previous research, yet there are still many more questions that are not known. For example, how do single components affect the experience and its ability to reduce stress? Should we add a lake, or would it enhance the sense of Biophilia more if the space were filled with more trees? VR allows ``magic'' to take place, yet we must empirically understand what works and does not.

This position paper will describe the main takeaways from multiple studies we have conducted on forest bathing in VR and identify areas where the community interested in this work may want to concentrate. 

      \begin{figure}[tbh]
        \center{\includegraphics[width=0.95\textwidth]
        {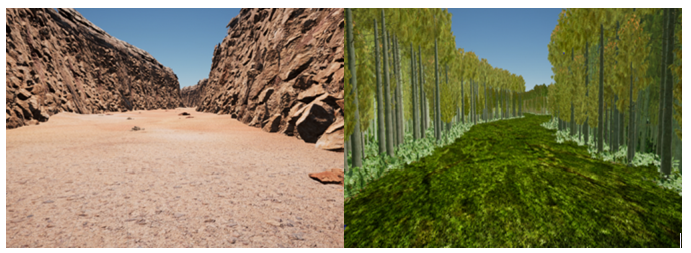}}
        \caption{\label{fig:canyonforest} Canyon and Forest Environments from ~\cite{masters22sap}}
          \Description{Two experimental VR environments (Canyon and Forest) from~\cite{masters22sap}}
      \end{figure}

\section{State of the Art}

The authors have participated in different studies showing the potential of VR forest bathing. In the first published paper, Masters et al. compared an environment without biomass (plant life) to one that only had plant life to investigate the role of biomass in the stress-reducing effects of nature~\cite{masters22sap}. A between-subjects experiment was conducted to compare a forest and canyon environment via three psychological measures: the Zuckerman Inventory of Personal Reactions (ZIPERS)~\cite{zuckerman1977development}, the Positive and Negative Affect Schedule (PANAS)~\cite{watson1988development}, and the Perceived Restorativeness Scale~\cite{hartig1996validation}. These measures were taken as a baseline when participants entered, after they completed the Markus and Peters Arithmetic Test~\cite{peters1998cardiovascular} to verify the efficacy of the stressor, and after they entered the environment for 10 minutes to compare the efficacy of the environments. Results indicated that the stressor was effective, but there were no significant results when comparing the two environments. In this particular stuyd, we only found non-significant trends suggesting that the forest environment may be more restorative, indicating the need for further research in this area. However, later studies below, there were some significant results, between the control condition and the forest environment~\cite{jnicoly2024sap} and between the control condition and the high-realism environment~\cite{masters24tap}.

In the second published paper, Masters et al. looked at the differences between low-fidelity and high-fidelity (i.e., low-realism versus high-realism) biomass~\cite{masters24tap}. One challenge with high-fidelity assets is that they are difficult to render in VR without causing lag and cybersickness. Since the goal of VR forest bathing is to make it accessible, it is necessary to understand the intricacies of asset design and whether low-fidelity assets that are feasible to deploy on consumer headsets can achieve similar stress-reducing effects. Masters et al. conducted a mixed-design study with a between-subjects comparison of a low and high-realism nature environment with a within-subjects control condition where participants quietly closed their eyes. Before all conditions, participants underwent the same stressor with the same psychological questionnaires as ~\cite{masters24tap}, as well as heart rate and blood pressure. All of these measurements were taken at three intervals per session, once as a baseline, then post-stressor, and then post-environment, to observe the stress-reduction capabilities of the environments. Overall, Masters et al. observed significant differences indicating that the stressor was effective but only found a significant difference in the PRS General Restorativeness category, meaning that participants in the high-realism environment perceived that they were more restored than those in the low-realism environment. However, trends in the rest of the data lean towards stress-reduction potential in the high-realism environment, and Masters et al. concluded their study by calling for future work to investigate this trend more deeply.

      \begin{figure}[tbh]
        \center{\includegraphics[width=0.95\textwidth]
        {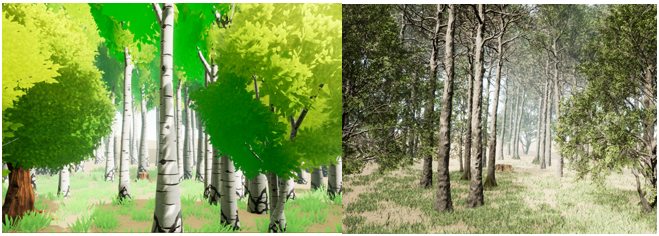}}
        \caption{\label{fig:highlow} Low and High Realism Environments from ~\cite{masters24tap}}
         \Description{Two experimental VR environments (Low and High Realism) from~\cite{masters24tap}}
      \end{figure}

In the third published paper, Nicoly et al. were interested in whether the benefits of VR forest bathing stem from being immersed in realistic, moving nature or from seeing something beautiful~\cite{jnicoly2024sap}. To observe this, three environments were designed: an abstract art environment that captured the beauty of art without any natural qualities, a real-life forest image that captured the full realism of real nature but without movement as it was an image, and a 3D, VR replica of the forest image that had high-fidelity 3D assets with wind movement. In the experiment, the user sat in an identical VR cabin, and each condition was visible on the wall of the cabin that the user's chair was facing. The three environments were compared in a between-subjects experiment along with the same control condition, stressor, measurements, and measurement intervals from ~\cite{masters24tap}. Nicoly et al. found the stressor effective, significant differences in ZIPERS positive affect and PRS General restorativeness between the 3D forest and control, and significant differences PRS between the 3D forest and abstract art. These results indicated that the realistic, moving nature environment shows some benefit over the art and control, but more work is needed to understand the benefits and optimal environment design.

We recently completed an experiment. Our initial findings suggest that older adults who use VR forest bathing improve their executive functions.

\section{Aging and VR nature}
The trajectory of cognitive impairment is thought to vary across types of dementia. However, one of the first and most critical types of cognitive decline thought to accompany ADRD is executive functioning (EF)~\cite{salat2001selective}. Executive functioning is an umbrella term encompassing a variety of higher-level cognitive processes that generally require attention, problem-solving, and mental flexibility. Of particular interest to researchers and clinicians is the relationship between EF and one’s quality of life in aging. Executive functioning is strongly related to one’s performance in activities of daily living and is the best predictor of functional status and functional deterioration in the case of pathological aging~\cite{cahn2000prediction}. Therefore, research has focused on interventions that target executive functioning both to sustain normal aging and ameliorate cognitive decline. Interventions focused on physical exercise have flourished, with findings suggesting that almost all types of physical exercise are associated with improved EF with those that are varied with diverse motor and diverse neuromuscular coordination having the greatest impact~\cite{cai2024effects}. There is evidence that suggests that this relationship is not merely correlational, interventions (e.g., exercise) that improve EF in ADRD patients also alleviate symptoms of cognitive decline~\cite{law2020physical}. Therefore, interventions that can improve EF are key targets for ADRD prevention.  

While interventions based on exercise are strongly connected to improvements in EF, there are limitations in terms of baseline physical health that is needed to engage in these types of activities. An alternative option that has gained traction in the past decade is exposure to nature. Attention Restoration Theory and Stress Recovery Theory have long suggested that simply spending time in nature can improve EF and improve stress recovery, respectively,~\cite{kaplan1995restorative, ulrich1981natural} with accumulating evidence in younger adults supporting these claims~\cite{cheng2021systematic, ohly2016attention, stevenson2018attention}. For example, recent meta-analyses on younger adult populations have found that acute (less than 1 hour) exposure to nature improves three core EF abilities: working memory, cognitive flexibility, and inhibition, in addition to improving mood/affect~\cite{ohly2016attention, stevenson2018attention, chang2023relationships}. While research focused on nature exposure and older adults is not as substantial as the literature on young adults, the benefits that have been observed are meaningful. One particular type of nature intervention, forest bathing, appears well-suited for older adults.  Forest bathing refers to spending time in nature while one engages all senses. Unlike exercise interventions, forest bathing is possible for those individuals who may not be suited for more strenuous interventions.  Current research suggests that forest bathing in older adults is associated with a variety of benefits, quality of life, and cerebral activity including endorsing a greater sense of purpose in life~\cite{chang2023relationships, garibay2024effects, mathias2020forest}. While forest bathing can be more inclusive than other types of interventions for older adults that require a degree of physical health and mobility, there are still limitations regarding access to nature, transportation to nature, and ability to tolerate weather issues. To address these issues, virtual-reality paradigms represent an exciting development that could extend the benefits of forest bathing to older adults who cannot engage in traditional forest bathing experiences.

VR techniques with older adults are an established method of providing experiences and interventions to promote well-being~\cite{appel2021virtual}. For example, music interventions have been successfully converted into VR format~\cite{faw2021being} for dementia patients. There is even some evidence that VR nature interventions can improve self-reported outcomes in patients with dementia~\cite{reynolds2018can, rose2021bringing}. However, to our knowledge, little work has directly examined whether VR nature can improve EF in healthy older adults, before disease onset. 

Our current pilot work addresses this gap directly, exposing older adults (aged 55+) to either VR nature, real-world nature, or a do-nothing control for 20 minutes. We find preliminary evidence that engaging with VR nature improves EF abilities in older adults above and beyond improvements seen in real-world nature or the control condition. While these findings are preliminary, they provide promising potential for VR to advance well-being in older adults. 

\section{Generative Forest Bathing}
The use of generative AI within VR extends the presented concepts further by enabling highly personalized and dynamic experiences~\cite{Gottsacker2024}. 
This technology can curate virtual transitions
~\cite{Feld2024} that replicate the calming effects of a forested environment, blending elements like forests, lakes, and other natural features tailored to individual needs. 
These AI-driven transitions could potentially enhance the therapeutic effects of virtual forest bathing by simulating the subtle, incidental experiences—such as the gradual transition from a dense forest to an open lakeside—that contribute to emotional and cognitive restoration.
Yet, this approach also introduces new challenges, particularly in understanding how these personalized and varied experiences influence the overall effectiveness of VR as a substitute for real nature. The complexity of AI-generated environments, where the content is tailored yet potentially inconsistent across users, requires careful empirical investigation to determine the true potential and limitations of virtual forest bathing as a stress-relief tool.

\section{Conclusion: A Vision Towards Positive Technology}

While recent findings have shown trends or findings (in some cases from the user's point of view, i.e., subjective), multiple questions have not been addressed. We are calling interested researchers to work together towards a common goal: to reduce stress and improve cognitive abilities, among other objectives, using Virtual Reality Forest Bathing. 

While the vision is clear—one that will provide equal access to reducing stress and improving cognitive functions using VR forest bathing—multiple questions need to be answered and work completed to achieve this goal. It is imperative that scientists drive future applications solidified in evidence that improve people's well-being. As mentioned, many questions have not been answered or need further proof to generalize them. From how cognitive functions can be enhanced using forest bathing, how to design the environments, and how to use generative AI to improve those designs adapting to the users' needs are still not entirely known. However, we believe the future is brighter if we use positive technology anchored in empirical and verifiable evidence.


\bibliographystyle{ACM-Reference-Format}
\bibliography{forest}

\end{document}